# Topological Hybrid Silicon Microlasers

Han Zhao[1], Pei Miao[1], Mohammad H. Teimourpour[2], Simon Malzard[3], Ramy El-Ganainy[2], Henning Schomerus[3] and Liang Feng[1*]

[1]*Department of Materials Science and Engineering, University of Pennsylvania, Philadelphia, PA 19104, USA*

[2]*Department of Physics and Henes Center for Quantum Phenomena, Michigan Technological University, Houghton, MI 49931, USA*

[3]*Department of Physics, Lancaster University, Lancaster, LA1 4YB, United Kingdom*

*Emails: fenglia@seas.upenn.edu

Topological photonics provides a robust framework for strategically controlling light confinement and propagation dynamics. By exploiting the marriage between this notion and symmetry-constrained mode competition in an active setting, we experimentally demonstrate, for the first time, a hybrid silicon microlaser structure supporting a topologically protected zero-mode lasing. This mode is distributed over an integrated array of optical microrings, each supporting multiple modes that compete for gain. However, the mode competition is diminished in favor of a pre-defined topological state, which reflects the charge conjugation symmetry induced by the gain profile. Robust single-mode operation in the desired state prevails even with intentionally introduced perturbations. The demonstrated microlaser is hybrid-implemented on a silicon-on-insulator substrate, thereby readily suitable for integrated silicon photonics with potential applications in optical communication and computing.

The discovery of topological band theory has ushered in a new era in condensed matter physics, providing intriguing insights into the world of low-dimensional quantum systems featuring e.g. the quantum Hall effect and quasiparticles with fractional statistics, and paving the way for engineering new states of matter, such as topological insulators and superconductors [1]. Inspired by this groundbreaking work, topological mechanisms of optical mode formation have been proposed by Haldane and Raghu [2]. The topological explorations in passive photonic systems have facilitated the design of structures with unidirectional transport channels in photonic crystals [3], optical cavity arrays [4,5] as well as helical waveguide lattices [6], and the demonstration of robust topological defect states in metamaterial arrangements and dielectric resonators [7,8] among a host of other intriguing effects [9–12]. On the other hand, the advent of parity-time-symmetric (and in general non-Hermitian) optical experiments [13–23] has widened the scope by the incorporation of gain and loss, which opens up avenues to explore new phenomena with no electronic analogues. These developments necessitated a reconsideration of the basic notions of topological protection in order to take into account the expanded design parameters space [24–31], and established a connection between topological physics and various separate activities on non-Hermitian photonic systems [32–34].

Active optical systems involving feedback mechanisms provide an even wider arena in which topological robustness collides with considerations of nonlinear dynamics. This poses many unexplored fundamental questions about the interplay between topological features, non-Hermiticity and the break-down of superposition principle. Here we explore the utility of topological concepts to active systems and experimentally demonstrate, for the first time, an on-chip hybrid silicon microlaser whose mode competition naturally favors a state arising from a topological defect. The topological laser structure considered here is motivated by a non-Hermitian variant [24,25] of the paradigmatic Su-Schrieffer-Heeger (SSH) model [35,36], a tight-binding model whose topological features arise from a sequence of alternating couplings and adapts flexibly to many physical settings [37,38]. These are realized using a microresonator array where the coupling profile is precisely controlled by their separation that in turn determines the strength of the evanescent wave tunneling rate (Fig. 1 (a)). On the other hand, the distributed gain and loss are provided by the pump beam and material engineering. The topological defect at the center of the array creates a topological zero-mode that decays exponentially away from the defect, and only populates one of the two sublattices. Spectrally, the defect state resides at the center of a band gap, where the symmetry of the passive band structure arises from a chiral symmetry [36]. The distributed gain and loss respect a non-Hermitian charge-conjugation symmetry (a non-Hermitian variant of the chiral symmetry), which results in a response that discriminates between the topological and non-topological states [25]. Considered in the complex frequency plane, this directly translates into an enhanced gain of the topological zero-mode state, therefore favoring it over other states throughout the nonlinear mode competition process (Fig.1 (b)).

Figure 1 (c) depicts an SEM picture of the fabricated topological microlaser array on a hybrid InGaAsP-silicon platform. It consists of nine coupled InGaAsP-silicon microring resonators on a silica substrate. The inner and outer ring radii are 3.5 $\mu m$ and 4.5 $\mu m$ with an estimated quality factor of ~4000, while the height of the silicon and InGaAsP layers are $220\ nm$ and $500\ nm$, correspondingly. The alternating edge-to-edge separations between the resonators are $d_1 = 300\ nm$ and $d_2 = 200\ nm$, which correspond to experimentally estimated alternating weak and strong couplings of $t_w = 78\ GHz$ and $t_s = 134\ GHz$. In order to generate the desired loss profile, a 10 $nm$ layer of Chromium (Cr) is deposited on top of every second resonator. Finally, the gain profile is provided though a uniform optical pumping applied from the top.

Figures 2 (a) show plots of the laser output intensity for three different pump levels. For a pump power of 0.71 GWm$^{-2}$ below the threshold, the spectrum of the light emission is broad and noisy. Right above the threshold, at 0.81 GWm$^{-2}$, a well-isolated resonant peak appears around a wavelength of $\lambda \sim 1523\ nm$. As the pump level is further increased to 1.84 GWm$^{-2}$, the single mode operation persists and the signal-to-noise ratio between the emission peak and the background noise increases. Figure 2 (d) shows the light-light plot, where the pump dependence of the total emitted intensity agrees well with the expectations for a single-mode laser as it only displays a single threshold without further kinks.

In order to assess the role of the topological features in the mode selection mechanism, we also present experimental data for the laser array before and after deposition of the Cr layer, representing uniform vs PT-type pumping, respectively. As shown in Fig. 3 (a) at a pump power of 1.61 GWm$^{-2}$, the uniform pumping scenario displays a broader emission spectrum with multiple peaks and a reduced peak intensity. Concurrently, the output emission from the homogenous device is distributed over the whole structure, and the internal emission patterns in each resonator display a reduced contrast. On the other hand, when the topological mode selection is enforced via the lossy Cr material, the emitted light is confined only to a narrow spectral peak as shown in Fig. 3 (b). The broadening of the spectrum in Fig. 3 (a) cannot be explained by the frequency splitting due to supermode formation alone (in the order of ~3 $nm$ in our system). It can be rather attributed to the mode competition between the different transverse modes in each resonator. The couplings between any pair of these modes belonging to adjacent resonators vary depending on the field overlap integrals, with TM$_{11}$ mode exhibiting the smallest coupling coefficients. Thus, even though each mode family forms its own collective topological state, adding the Cr layer spoils the quality factor of these modes unevenly, eventually favouring the defect state made from the hybridization of the TM$_{11}$ modes.

One of the most important features of topological defect states is their robustness against disorder. Particularly, the spectral features of the zero-mode associated with the passive SSH model are known to be insensitive to off-diagonal perturbations represented by the coupling coefficients. In active SSH systems, however, the system can still display a certain level of immunity against diagonal perturbations, which can introduce a shift in the emitted frequency without supressing lasing of the zero-mode defect state. This behaviour is exemplified in Figs. 4 (a) depict the modelled spectrum (top panel) and defect mode profile (lower panel) for the laser array when the resonant frequency of the third microring from the right is perturbed (as shown schematically in the top panel) by one part in thousand. To confirm these predictions experimentally, we introduced a polymer layer on top of the third ring resonator from the right to introduce a shift in its resonant frequency (see SEM image in the top panel of Fig.4 (b)) and measured the emission spectrum and profile as depicted in the middle and lower panels of Figs. 4 (b). Evidently, the defect state lasing with a high signal to noise ratio still persist, without appreciable change in the spatial emission profile apart from very small intensities leaking to the otherwise dark sublattice.

In conclusion, we have presented the first demonstration of topological zero-mode lasing in an SSH laser array implemented in a hybrid silicon platform. Our experimental work, supported by theoretical and numerical analysis, opens new doors for exploring the interplay between topology and non-Hermiticity in nonlinear setups, and provides an avenue for utilizing these ingredients to engineer and study new phenomena with no electronic analogue. From a practical perspective, our work demonstrates the ability to employ topological features in order to control the collective dynamical behaviour of laser arrays by enhancing the lasing of the topological defect state and supressing the extended states by spoiling their quality factors. This feature may prove useful in future work for engineering laser arrays that exhibit single supermode operation at high power levels, which is currently one of the major challenges in laser engineering. Owing to their topological features, such laser arrays would be immune to moderate perturbations such as heating-induced resonant shifts, whilst posing fewer constraints on the geometry and requiring less sophisticated electronics to ensure their proper operation. Equally important, our demonstrated device is implemented in hybrid-silicon platform, which lends itself naturally to integration with other components in silicon photonic circuits, a potential advantage for telecommunication applications.

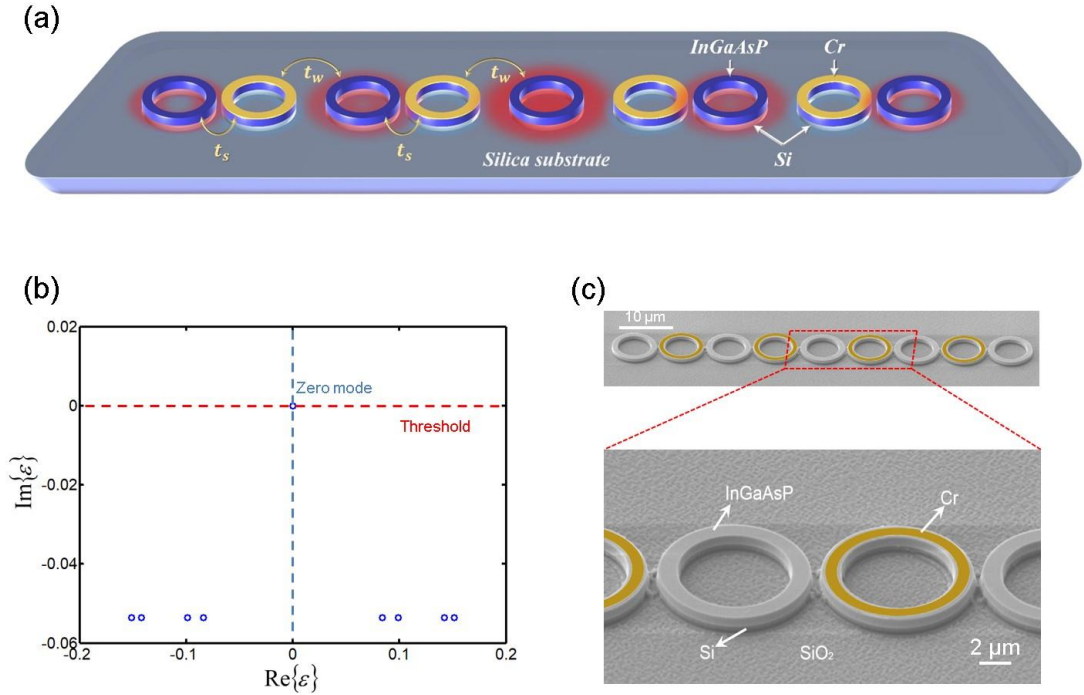

**Figure 1. Topological hybrid silicon microlaser.** (a) Schematic of a topological laser array made of nine microring resonators with alternating weak and strong coupling profile, emulating an SSH array. A layer of Cr (shown in yellow) is deposited on top of every second element to introduce distributed loss. The red halos represent the intensity profile of the defect state that resides at the central site and decays exponentially away from the center, with zero intensity in every second element. (b) Spectral features of the SSH laser array, highlighting the lasing selectivity of the topological zero-mode defect. Due to the charge conjugation symmetry, the spectrum is symmetric around the imaginary axis. (c) An SEM picture of the fabricated structure consisting of nine rings, each made of InGaAsP quantum well layers on top of a thin silicon layer grown on silica substrate. All the dimensions are listed in the text.

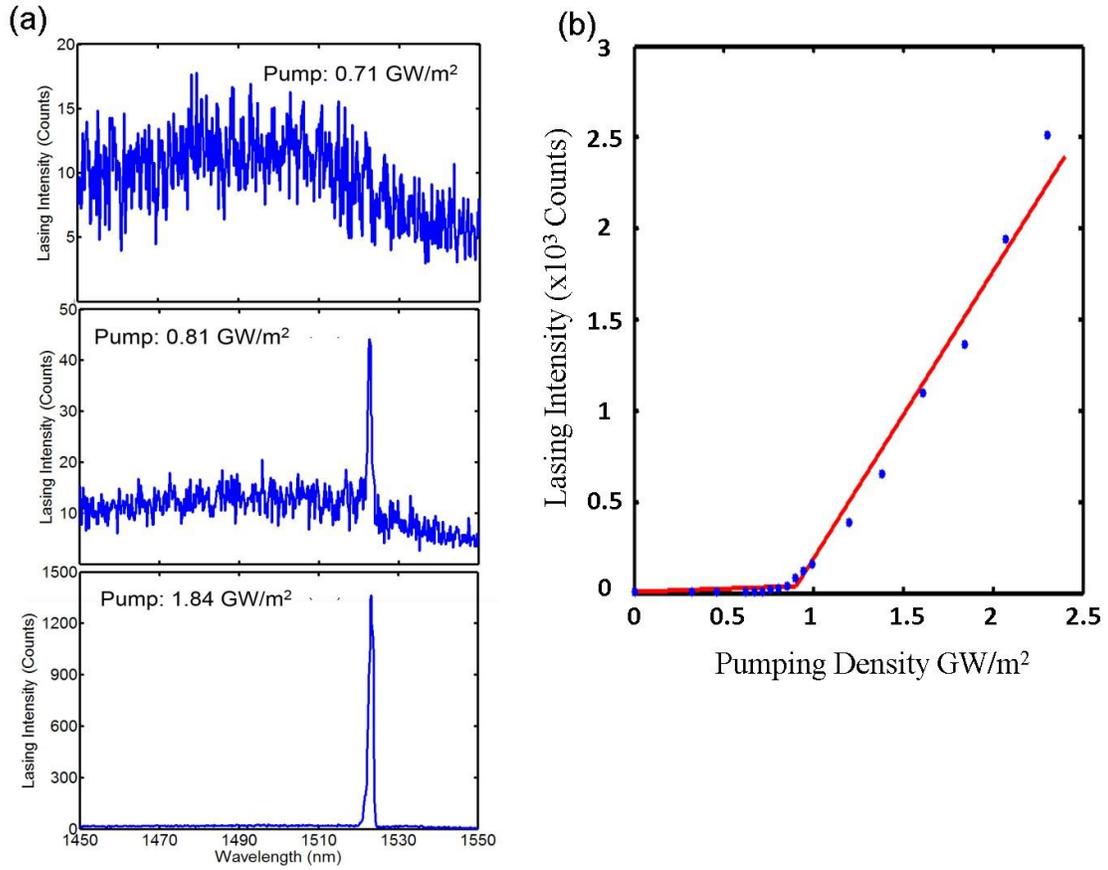

**Figure 2.** (a) Lasing spectrum of the structure of Fig.1 (c) as a function of the pumping power below threshold (top panel), just at threshold (middle panel) and well above threshold (bottom panel). As expected, as the pump power is increased, the output emission experiences a transition from broadband noisy fluorescence to a single narrow band peak with increased signal to noise ratio. (b) Pump dependence of the laser emission intensity, demonstrating the fingerprint of single mode lasing: only one threshold with no kinks in the light-light curve.

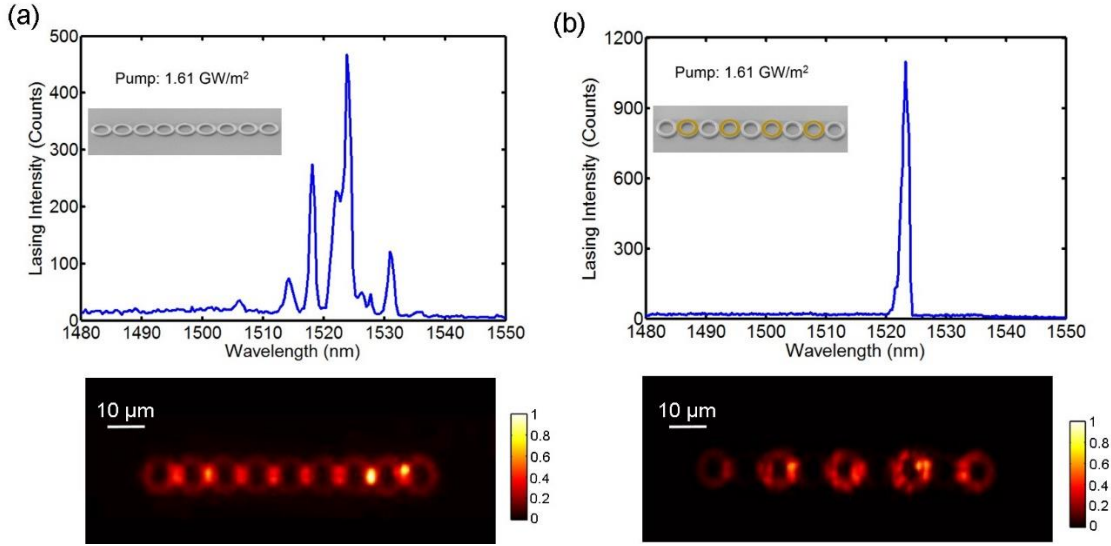

**Figure 3.** Emission spectra (top panels) and lasing mode profiles (lower panels) for SSH laser arrays under (a) Uniform pumping (before adding the Cr layer) and (b) PT-like pumping. The broad spectrum in (a) is due to mode competition between the transverse modes of the single ring resonators, each of which forming its own collective state. The addition of the lossy Cr layer judiciously spoils the quality factors of the strongly hybridized modes, favouring the topological state centred around $\lambda = 1523\ nm$.

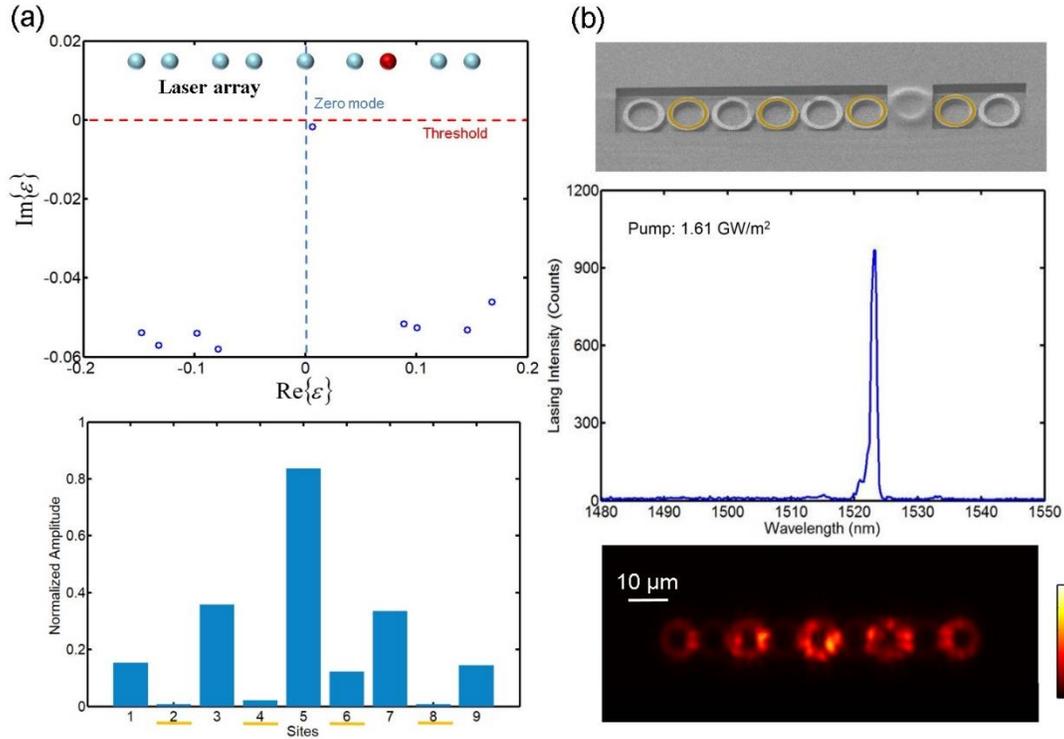

**Figure 4.** Effect of disorder on the lasing characteristics of the topological SSH laser. (a) Numerical calculations for the spectrum of a disordered lattice (top panel) when a perturbation is introduced to the third site from the right, as indicated schematically at the top of the same panel. The corresponding mode profile is shown in the lower panel, where the general characteristics of the zero-mode persist with addition of small light intensities on the previously dark sublattice. To confirm these general predictions we perturbed the laser array by depositing a layer of polymer on the corresponding ring, as shown in the SEM image in the top panel of (b). The middle and lower panels depict the emission spectrum and mode profile, respectively, where clear features of single mode operation are observed.